\documentclass[preprint,showpacs,aps]{revtex4}
\usepackage{graphicx}
\usepackage{epsfig}
\usepackage{CJK}
\usepackage{amsmath}
\usepackage{supertabular}
\usepackage{textcomp}

\begin{document}

\title{Investigation of the short-range correlation and its induced high-momentum tail by photon emission from intermediate energy heavy-ion reactions}
\author{Zhi Wang$\footnote{E-mail address: wangzhi.mail@foxmail.com}$}

\address{School of Mathematics and Statistics, Guizhou University of Finance and Economics, Guiyang 550025, China}

\noindent

\begin{abstract}

The effect of high-momentum tail (HMT) in nucleon momentum distribution of neutron-rich nuclei has been investigated by employing proton-neutron bremsstrahlung photons in the reactions of $^{124}$Sn +$ ^{124}$Sn, $^{112}$Sn + $^{112}$Sn and $^{197}$Au + $^{197}$Au at intermediate energies based on the IBUU transport model. With the time evolution of photon multiplicity and the double differential probability of photon production, we show the emission of bremsstrahlung photons is sensitive to the high-momentum component in nucleon momentum distribution of neutron-rich nuclei, and the HMT can lead to an obvious increase of hard photon production from proton-neutron bremsstrahlung collisions.
Finally, the ratio of double differential photon production probability is examined as possible probe of the HMT in nucleon momentum distribution of neutron-rich nuclei.

\end{abstract}

\pacs{25.70.-z, 24.10.Lx, 21.65.Cd}

\maketitle

\section{Introduction}
The study of the nucleon momentum distribution in nuclei or nuclear matter is of great interest due to its close relation to many important phenomena in both nuclear physics and hadronic physics \cite{1,2,3}. Containing the information of both the nuclear mean field properties and the short-range interaction of nucleon-nucleon pairs, a lot of efforts have recently been invested in studying the nucleon momentum distributions in symmetric or asymmetric nuclear systems from the perspective of theories \cite{4,5,7} and experiments \cite{8,9,10}. Various microscopic and phenomenological models indicate that the nucleon-nucleon short-range correlation (SRC) induced by tensor force pushes a part of nucleons in nuclei from low momenta ($k<k_F$ , where $k_F$ is the Fermi momentum of nucleon) to high momenta ($k>k_F$ ), and leads to a high-momentum tail (HMT) in single-nucleon momentum distribution \cite{11}.

Experimentally, there are several new progresses in the study of the nucleon momentum distribution and the SRC between nucleon-nucleon pairs in recent years \cite{12,15}. For example, the high-energy electron-scattering experiment on the nucleus $^{12}$C carried out by the Jefferson Lab has shown that about 20\% of nucleons in $^{12}$C are correlated, where 90\% of the correlated pairs are in the form of neutron-proton (n-p) SRC pairs, with the remaining 10\% being p-p and n-n pairs \cite{12}. This high-momentum part in the nucleon momentum distribution induced by SRC can lead to a larger average kinetic energy of nucleons as compared to the uncorrelated free Fermi gas (FFG) \cite{13,14}. More recently, the high-momentum transfer experiment using high-energy electron-scattering on $^{27}$Al, $^{56}$Fe, and $^{208}$Pb targets performed by the Jefferson Lab \cite{15} further suggests that even in heavy, isospin-rich nuclei, the nucleons short range interactions can form correlated high momentum pairs and the universal nature of SRC pairs in neutron-rich nuclei are still predominantly n-p correlations. Such phenomenon is interpreted by the existence of strongly tensor force in the n-p deuteron-like state \cite{16,17}.
Since this isospin-dependent SRC is stronger in n-p pairs with respect to n-n or p-p pairs, protons have greater probability to have momenta higher than the Fermi momentum, as compared with neutrons in isospin asymmetric nuclei \cite{15}.
Even in the neutron stars, due to the number of protons account for only a small fraction of nucleons, the above predominant n-p SRC will lead to average kinetic energy of protons far greater than that of neutrons \cite{18}.

The high-momentum component in nucleon momentum distribution induced by SRC are important for many issues involving nuclear structure and nuclear reaction \cite{19,20,21,22}. For instance, this isospin dependence of nucleon-nucleon SRC has an immediate impact on the nuclear symmetry energy \cite{20,21}. When taking the HMT of nucleon momentum distribution into consideration, the kinetic component of nuclear symmetry energy will be decreased substantially (even become negative value). This has been calculated in both microscopic many-body theories \cite{23,24,25} and phenomenological models \cite{13,26,27}.

It is clear that the high-momentum component of momentum distribution in nuclei can affect the yields of mesons (such as $\pi$) in intermediate energies heavy-ion reactions \cite{18}.
However, these probes can interact with the nuclear medium so that the information they carried may bring a blurred image of their source \cite{28}. To overcome this difficulty, the electromagnetic observable of energetic photons is proposed as an attractive alternative to explore the properties of nuclear matter.

During the past two decades, photon emission from heavy-ion reactions have been studied both experimentally and theoretically \cite{29,Grosse,Bauer,30,31,32,33,34}. The high-energy bremsstrahlung photons are found to be mainly produced in the early stage of the reactions, which interact with nucleons only electromagnetically and will escape almost freely from the nuclear environment once produced, so they have very low probability to experience final state interaction. This makes hard photons to be a clean sensitive probe in nucleus-nucleus collisions.
Moreover, many experimental facts and theoretical calculations \cite{28,35} have shown that in intermediate energy heavy-ion reactions the hard photons are mainly produced from incoherent proton-neutron bremsstrahlung collisions, namely, $pn\rightarrow$$pn\gamma$. The intensity of proton-proton bremsstrahlung is approximately an order of magnitude smaller than that of proton-neutron bremsstrahlung \cite{29,36}. Therefore it is expected that the proton-neutron bremsstrahlung photons can serve as a potential probe of HMT in nucleon momentum distribution.

The HMT in nucleon momentum distribution of symmetric nuclear system has recently been discussed by using the proton-neutron bremsstrahlung photons in the ${}^{12}$C + ${}^{12}$C reaction within the framework of the isospin-dependent Boltzmann-Uehling-Uhlenbeck (IBUU) transport model \cite{37}. It is found that the production of bremsstrahlung photons is significantly promoted when taking the HMT in nucleon momentum distribution of $^{12}$C into account. Then several interesting questions arise naturally: Whether the bremsstrahlung photons are sensitive to the high-momentum component of asymmetric nuclear systems? What is the difference of photon emission between FFG and HMT cases in asymmetric nuclear systems?
And can we use the photon emission from intermediate energy heavy-ion reactions as a probe of the HMT of neutron-rich nuclei?
In this work we will attempt to answer these questions by employing photon emission from proton-neutron bremsstrahlung collisions in the reactions of ${}^{124}$Sn + ${}^{124}$Sn, ${}^{112}$Sn + ${}^{112}$Sn and ${}^{197}$Au + ${}^{197}$Au at intermediate energies (such as 50 MeV/nucleon and 140 MeV/nucleon) based on the IBUU transport model. The photon multiplicity and double differential probability of photon production are discussed in detail by assuming two kinds of momentum distribution in the neutron-rich nuclei, namely, the one of FFG and another with HMT based on Refs. \cite{15,26}.
The paper is organized as follows. In next section, the theoretical framework of IBUU transport model is briefly reviewed and the calculation method of photon production is introduced, and the parameterizations of nucleon momentum distribution in asymmetric nuclear matter is also performed. The numerical results and discussions are present in section III. The summary is given in last section.
\section{Theoretical framework}
\subsection{The isospin-dependent Blotzmann-Uehling-Ulenbeck equation}
The isospin-dependent Boltzmann-Uehling-Uhlenbeck (IBUU) model \cite{38,39,40} has been used successfully in describing the dynamical evolution of nuclear collisions at intermediate energies. This model describes time evolution of the single particle phase space distribution function $f(\vec{r},\vec{p},t)$, the main equation of the IBUU model takes form as:
\begin{equation}
\frac{\partial f}{\partial t}+\vec{v} \cdot \nabla_{r}f-\nabla_{r} U \cdot \nabla_{p}f=I_{collision} .\tag{1}
\end{equation}
The distribution function $f(\vec{r},\vec{p},t)$ is interpreted as the probability of detecting a particle at position $\vec{r}$ with momentum $\vec{p}$ at a given time $t$. The left hand side of Eq. (1) means the dynamics evolution of the particle phase space distribution function due to nucleus mean field and its transport, where the mean-field potential $U$ depends on position and momentum of the particle and can be computed self-consistently using the phase space distribution function $f(\vec{r},\vec{p},t)$.
The collision item $I_{collision}$ on right hand side accounts for the modification of distribution function $f(\vec{r},\vec{p},t)$ by elastic and inelastic two body collisions with the Pauli blocking effect considered \cite{41}.

In the present work, we use two different kinds of mean-field potentials $U$ for simulation. One is the Skyrme-type parametrization for the soft Bertsch-Kruse-Das Gupta (SBKD) mean-field potential, which reads \cite{42}
\begin{equation}
U(\rho)=A(\rho/\rho_{0})+B(\rho/\rho_{0})^{\sigma},\tag{2}
\end{equation}
where $\sigma=7/6$ and $\rho_{0}=0.16$ fm$^{-3}$ is the saturation density, $A=-356$ MeV accounts for the attractive part, and $B=303$ MeV accounts for the repulsive part. This corresponds to the ground-state compressibility coefficient of nuclear matter $K=201$ MeV. The other is the extensively used isospin- and momentum-dependent baryon mean-field potential (MDI) as follows \cite{43}:
\begin{equation}
\begin{aligned}
U(\rho,\delta,\vec{p},\tau)&=A_{u}(x)\frac{\rho_{\tau'}}{\rho_{0}}+A_{l}(x)\frac{\rho_{\tau}}{\rho_{0}}          \nonumber    \\
                            &+B(\frac{\rho}{\rho_{0}})^{\sigma}(1-x\delta^{2})-8x\tau \frac{B}{\sigma+1} \frac{\rho^{\sigma-1}}{\rho_{0}^{\sigma}} \delta \rho_{\tau'}  \nonumber \\
                            &+\frac{2C_{\tau,\tau}}{\rho_{0}}\int d^{3}\vec{p'} \frac{f_{\tau}(\vec{r},\vec{p'})}{1+(\vec{p}-\vec{p'})^{2}/\Lambda^{2}}              \nonumber \\
                            &+\frac{2C_{\tau,\tau'}}{\rho_{0}}\int d^{3}\vec{p'} \frac{f_{\tau'}(\vec{r},\vec{p'})}{1+(\vec{p}-\vec{p'})^{2}/\Lambda^{2}},
\end{aligned}\tag{3}
\end{equation}
where $\delta=(\rho_n-\rho_p)/(\rho_n+\rho_p)$ is the isospin asymmetry, $\rho=\rho_n+\rho_p$ is the nucleon density, and $\rho_n, \rho_p$ are the neutron and proton densities, respectively. $\tau = 1/2 (-1/2)$ for the neutron (proton) and $\tau\neq\tau'$.
The variable $x$ is used to mimic different forms of the symmetry energy.

In the IBUU model, the isospin-dependent nucleon-nucleon (NN) scattering cross section is also an important input quantity. Here we adopt the isospin-dependent in-medium NN scattering cross section from the scaling model according to nucleon effective mass \cite{40,44}, which is given by:
\begin{equation}
\sigma_{NN}^{medium}=\sigma_{NN}^{free} \cdot (\mu_{NN}^{\ast}/\mu_{NN})^{2},\tag{4}
\end{equation}
where $\mu_{NN}$ and $\mu_{NN}^{\ast}$ are the reduced masses of the colliding nucleon-pairs in free space and nuclear medium, respectively. $\sigma_{NN}^{free}$ is the free-space NN scattering cross section taken from experimental data \cite{33}.

\subsection{Production cross sections of bremsstrahlung photons}

Due to the photon production probability is so small that only one photon produced roughly in one thousand nucleon-nucleon collisions, thus the dynamical evolution of photon production probability in intermediate energy nuclear reactions can be calculated by a perturbative approach \cite{45}. With this method one can calculate the photon production probability at each proton-neutron collision and the total production probability can be obtained by summing over all proton-neutron collisions in the entire history of the reaction \cite{32,33}.

For hard photon production from proton-neutron bremsstrahlung, we use the prediction of the one boson exchange model involving the quantum effects, in which the expression of the double differential photon production probability is given by Gan et al. \cite{46} as:
\begin{equation}
\frac{d^{2}P}{d\Omega dE_{\gamma}}=1.67\times10^{-7}\frac{(1-y^{2})^{\alpha}}{y},\tag{5a}
\end{equation}
where $y=E_{\gamma}/E_{max}$, $E_{\gamma}$ is energy of the produced photon and $E_{max}$ is the energy available in the center-of-mass of the colliding proton-neutron pairs. $\alpha=0.7319-0.5898\beta_{i}$, $\beta_{i}$ is the initial velocity of the proton in the proton-neutron center-of-mass system.

There is also another probability formula which is taken from the semiclassical hard sphere collision model has been widely used in the literatures \cite{41,45}. The double differential probability, ignoring the Pauli exclusion in the final state, is given by
\begin{equation}
\frac{d^{2}P'}{d\Omega dE_{\gamma}}=6.16\times10^{-5}\times\frac{1}{E_{\gamma}}(3\sin^{2}\theta_{\gamma}\beta_{i}^{2}+2\beta_{f}^{2}),\tag{5b}
\end{equation}
where $\theta_{\gamma}$ is the angle between incident proton direction and emission direction of photons; and $\beta_{i}$ and $\beta_{f}$ are the initial and final velocities of the proton in the proton-neutron center of mass frame, respectively.

\subsection{Parameterization of the nucleon momentum distribution}
In the IBUU transport model, the initial phase-space function is dependent on the nucleon density and momentum distributions in both projectile and target nuclei. The isospin-dependent proton and neutron density distributions of the initial colliding nuclei are decided by the Skyrme-Hartree-Fock with Skyrme $M^\ast$ (SM) force parameter \cite{47}. As to the nucleon momentum distributions in initial collision nuclei,
here we introduce two versions of initial single-nucleon momentum distributions in asymmetric nuclear matter (ANM) for IBUU calculations. One is the so-called free Fermi gas (FFG) distribution, in which the momentum distribution is simply a step function of a constant for $k\leq k_F$, and zero for $k>k_F$ where $k_F$ is the nucleon Fermi momentum. The other is the single-nucleon momentum distribution with a high-momentum tail (HMT) as shown below.

The proton and neutron momentum distributions with HMT of ANM are proposed based on the following considerations: (i) the protons have larger probability than neutrons to have momenta higher than the Fermi momentum due to the n-p SRC in neutron-rich nuclei \cite{15,M.M}, (ii) the recent theories and experiments have suggested that the HMT of nuclear momentum distribution decreases as $C/k^4$ \cite{12,15,26}, and (iii) the approximately linear-relationship between high-momentum component of nuclear momentum distribution and isospin asymmetry are presented by microscopic approaches and phenomenological models \cite{48,49,50}. Thus we parameterize the $n(k)$ of nucleons in ANM as:
\begin{equation}
{n_{ANM}}{(k)_p} = \left\{ {\begin{array}{*{20}{c}}
{{C_1}{\kern 1pt} }\\
{{{{C_2}(1 + \delta )} \mathord{\left/
 {\vphantom {{{C_2}(1 + \delta )} {{k^4}{\kern 1pt} }}} \right.
 \kern-\nulldelimiterspace} {{k^4}{\kern 1pt} }}}\\
0
\end{array}} \right.\begin{array}{*{20}{c}}
{(k \le k_F^p)}\\
{(k_F^p < k \le \lambda k_F^p)}\\
{(k > \lambda k_F^p)}
\end{array},\tag{6}
\end{equation}
\begin{equation}
{n_{ANM}}{(k)_n} = \left\{ {\begin{array}{*{20}{c}}
{{{C'}_1}{\kern 1pt} {\kern 1pt} }\\
{{{{C_2}(1 - \delta )} \mathord{\left/
 {\vphantom {{{C_2}(1 - \delta )} {{k^4}{\kern 1pt} }}} \right.
 \kern-\nulldelimiterspace} {{k^4}{\kern 1pt} }}}\\
0
\end{array}} \right.\begin{array}{*{20}{c}}
{(k \le k_F^n)}\\
{(k_F^n < k \le \lambda k_F^n)}\\
{(k > \lambda k_F^n)}
\end{array},\tag{7}
\end{equation}
where $\delta$ is the isospin asymmetry, $k_{F}^{p(n)}=[3\pi^{2}\rho_{p(n)}]^{\frac{1}{3}}$ is the nucleon Fermi momentum of proton (neutron) and parameter $\lambda \approx 2.75 $ is the high-momentum cutoff \cite{26}. The parameters $C_{1}$ and $C_{1}^\prime$ are determined by the normalization condition of proton and neutron momentum distributions:
\begin{equation}
4\pi\int_{0}^{\infty} n_{ANM}(k)_{p(n)} k^{2}dk=1.\tag{8}
\end{equation}
The parameter $C_{2}$ leads to about 25\% nucleons of symmetric nuclear system ($\delta=0$) involved in the HMT \cite{26}:
\begin{equation}
4\pi\int_{k_{F}^{p(n)}}^{\lambda k_{F}^{p(n)}} n_{ANM}(k)_{p(n)} k^{2}dk=0.25.\tag{9}
\end{equation}

Based on the above nucleon momentum distribution with HMT, one can obtain that there are roughly $25\%\times(1+\delta)$ protons ($25\%\times(1-\delta)$ neutrons) with momenta larger than the proton (neutron) free Fermi momentum.
Note that one can obtain the single-nucleon momentum distribution function with HMT in symmetric nuclear systems such as $^{12}$C when setting the isospin asymmetry $\delta=0$ in the Eqs. (6) and (7).
According to the local density approximation \cite{51}, the nucleon momentum distribution in finite nucleus can be calculated from the Eqs. (6) and (7):
\begin{equation}
n_{p(n)}(k)=\int d^{3}r \, \rho_{p(n)}(r) \, n_{ANM}(k,k_{F}^{p(n)}(r)),\tag{10}
\end{equation}
where $n_{p(n)}(k)$ is the nucleon momentum distribution in finite nucleus such as $^{124}$Sn  \cite{wang}. $\rho_{p(n)}(r)$ is the proton (neutron) density distribution of initial nucleus, and $n_{ANM}(k,k_{F}^{p(n)}(r))$ depicts the proton (neutron) momentum distribution in ANM.

\section{numerical results and discussions}

In this work, we have mainly simulated ${}^{124}$Sn + ${}^{124}$Sn reaction at beam energies of 50 MeV/nucleon and 140 MeV/nucleon within the IBUU model, in which two different versions of nucleon momentum distribution, \emph{i.e.}, the HMT and FFG cases, are introduced. We use the bremsstrahlung photons, especially the high-energy bremsstrahlung photons, as potential observable to investigate the effect of HMT in neutron-rich nuclei.
First, we would like to
compare the result of theoretical calculation based on the IBUU model with available experimental data.
We find that there exist experimental data on the photon production in the reaction of ${}^{12}$C + ${}^{12}$C in the literature \cite{Grosse}.
To compare the experimental data, we simulate the reactions of ${}^{12}$C + ${}^{12}$C at three different beam energies of 60, 74 and 84 MeV/nucleon.

\begin{figure}[htb]
\centering
\includegraphics[width=0.8\linewidth,angle=0,clip=true]{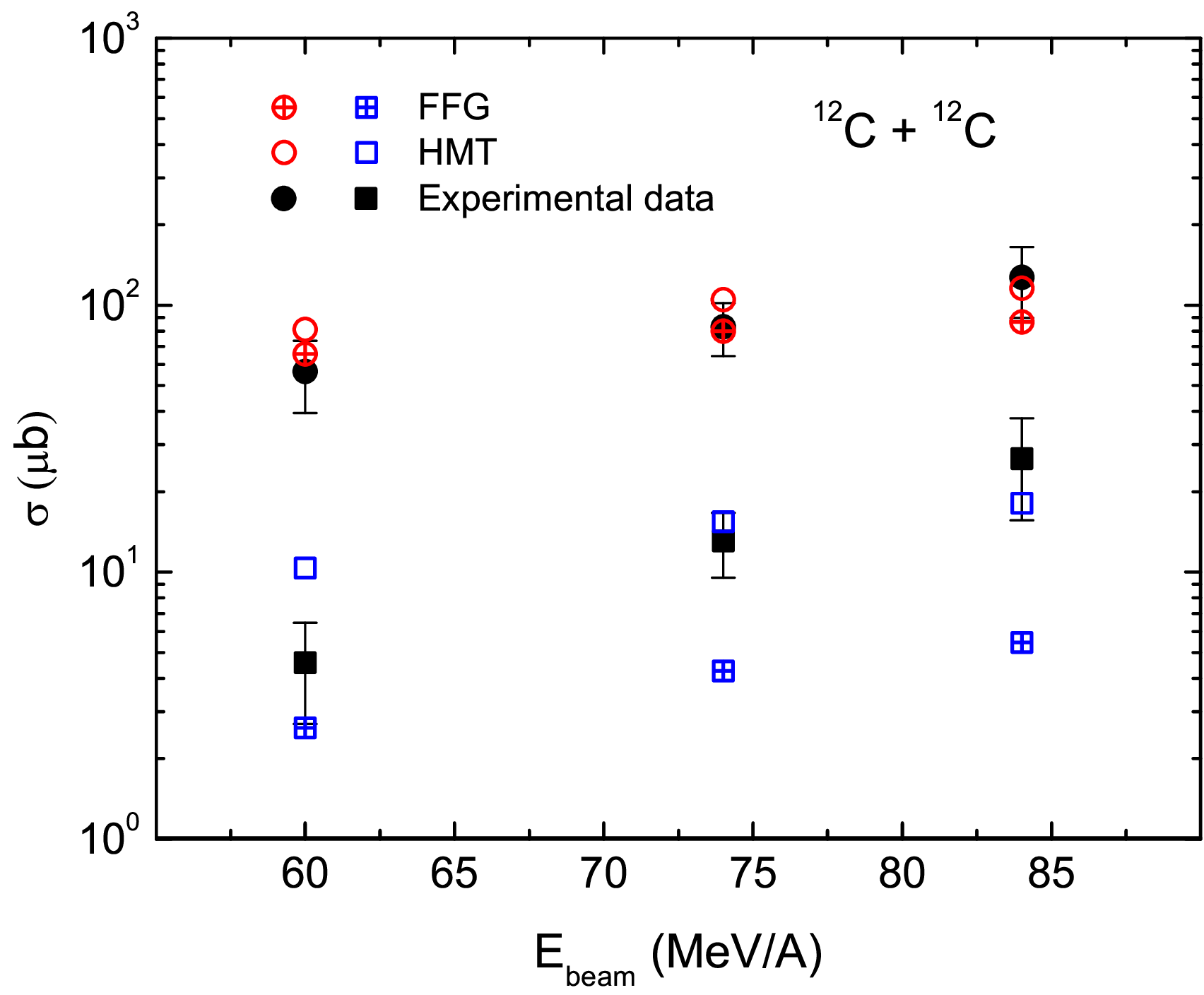}
\caption{Beam energy dependence of the inclusive photon production cross sections in ${}^{12}$C + ${}^{12}$C collisions. The cross center symbols are the FFG case, the open symbols are the HMT case and the solid symbols are from experimental data \cite{Grosse}(The circles are for energy region 50 MeV $\leq E_{\gamma} \leq$ 100 MeV and squares for 100 MeV $\leq E_{\gamma} \leq$ 150 MeV).}
\end{figure}
In Fig. 1, we show the comparison between the calculated results in both FFG and HMT cases and the experimental data for the inclusive cross sections of hard photon production in the reaction of ${}^{12}$C + ${}^{12}$C.
Comparing the case of FFG, we notice that the HMT obviously increases the high energy photon production cross section.
It is seen that the calculated photon production cross sections in the FFG and HMT cases
are both in reasonable agreement with the experimental data for the hard photons with energy 50 MeV $\leq E_{\gamma} \leq$ 100 MeV.
However, for the energetic photons with energy 100 MeV $\leq E_{\gamma} \leq$ 150 MeV, the case of HMT fits the experiment date better.

\begin{figure}[htb]
\centering
\includegraphics[width=0.8\linewidth,angle=0,clip=true]{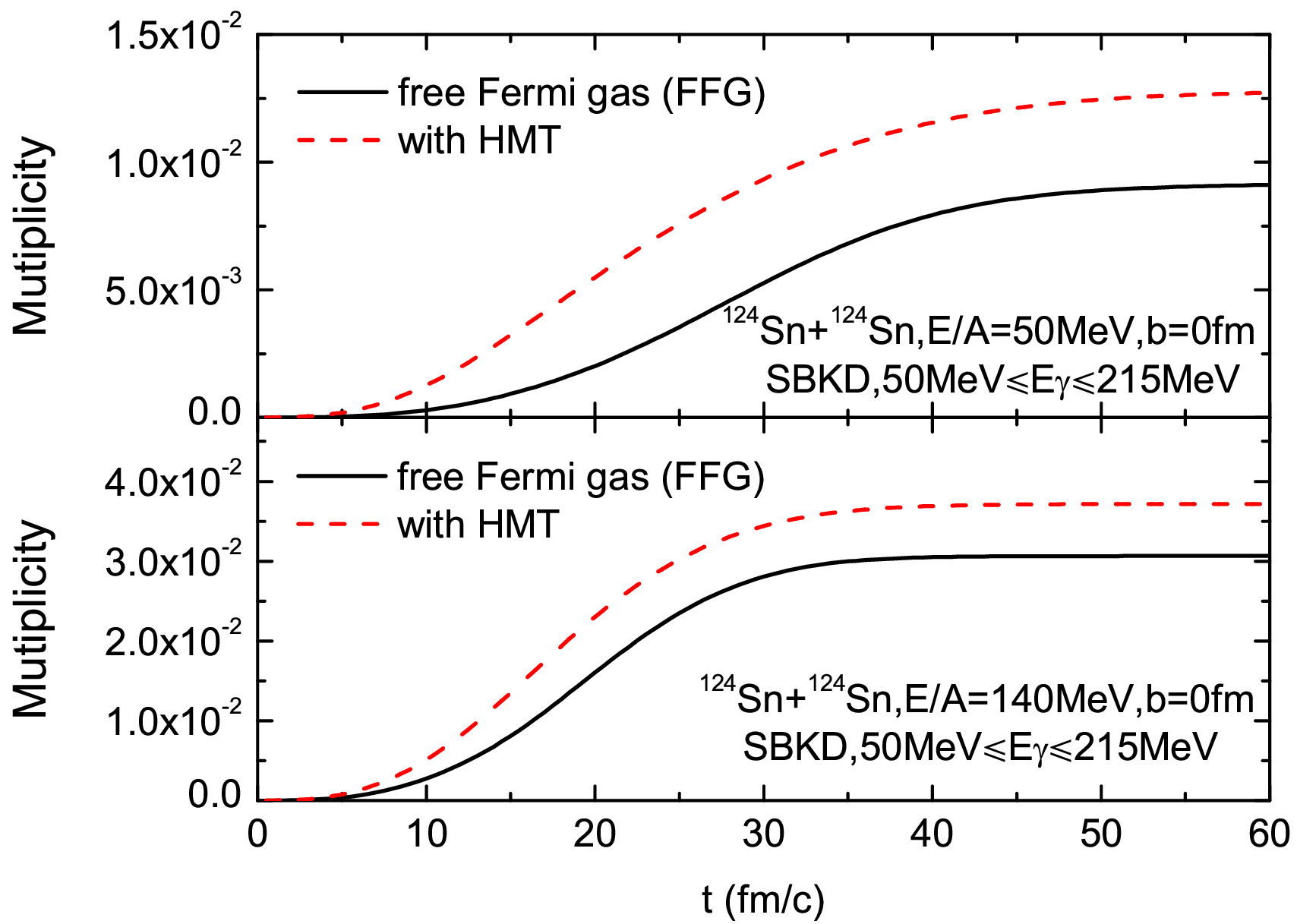}
\caption{Time evolution of multiplicity of hard photons with energy 50 MeV $\leq E_{\gamma} \leq$ 215 MeV in ${}^{124}$Sn + ${}^{124}$Sn central reaction in the FFG and HMT cases, respectively.}
\end{figure}
Shown in Fig. 2 is the time evolution of multiplicity of bremsstrahlung photons within energy range 50 MeV $\leq E_{\gamma} \leq$ 215 MeV in the head-on collisions at beam energies of 50 MeV/nucleon (upper panel) and 140 MeV/nucleon (lower panel), where the red dashed curve and black solid curve stand for the calculation results of HMT and FFG situations, respectively.
The multiplicity is defined as the integration of double differential probability of bremsstrahlung photons $d^{2}P/d\Omega dE_{\gamma}$ with respect to photon energy $E_{\gamma}$ and solid angle $\Omega$.
We see that bremsstrahlung photons are mainly produced in the early stage of heavy-ion collisions, for example the photon multiplicity become stable after about 40 fm/c at a beam energy of 140 MeV/nucleon reaction. It is also clear that at both the two beam energies, the HMT can lead to the obvious increase of high-energy bremsstrahlung photons compared with FFG case. This is quite natural that as the HMT in nucleon momentum distribution increases the average nucleon kinetic energy in both the projectile and target, then the yield of high-energy bremsstrahlung photons is promoted. With the growing of incident beam energy, the photon production is significantly increased. For instance, the multiplicity of bremsstrahlung photons in the FFG case increases from $9.1\times10^{-3}$ at 50 MeV/nucleon beam energy to $3.06\times10^{-2}$ at 140 MeV/nucleon case. However, the effect of HMT on the bremsstrahlung photons becomes weaker at higher incident beam energy. To be specific, there are approximately 37\% growth of the photon production for the reaction at incident beam energy of 50 MeV/nucleon while about 21\% growth of the photon production for the reaction at incident beam energy of 140 MeV/nucleon. This is understandable that as the beam energy increases, the initial movement of nucleons, including the SRC of nucleons in the nucleus becomes less important in nucleus-nucleus collisions.

\begin{figure}[htb]
\centering
\includegraphics[width=0.8\linewidth,angle=0,clip=true]{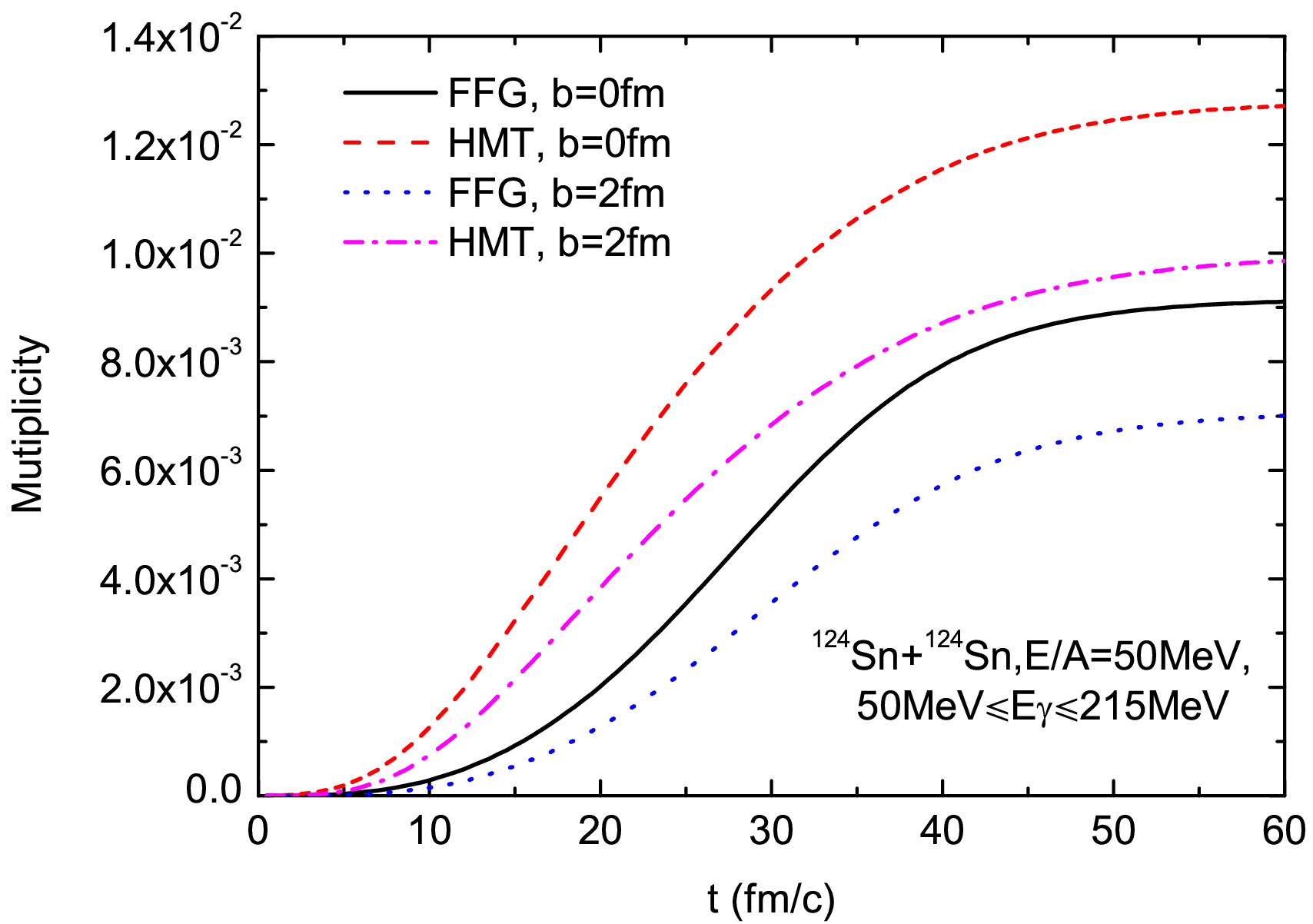}
\caption{Time evolution of the photon multiplicity in the ${}^{124}$Sn + ${}^{124}$Sn collisions at a beam energy of 50 MeV/nucleon with different impact parameter settings in both the FFG and HMT cases.}
\end{figure}
To investigate the influence of impact parameter on the photon production in the reaction, we plot in Fig. 3 the time evolution of photon multiplicity in the ${}^{124}$Sn + ${}^{124}$Sn reactions at a beam energy of 50 MeV/nucleon with different impact parameter $b$ settings. We see the effects of the HMT are very obvious and lead to about 30-40\% increasing of the photon production compared with FFG case at both the two impact parameters. It is also seen that the emission of hard photons is decreased with the growing of the impact parameter, for example, the multiplicity of bremsstrahlung photons in HMT case decreases from $1.27\times10^{-2}$ to $9.86\times10^{-3}$ when the impact parameter increases from 0 fm to 2 fm. This is simply because there are more nucleons participated in the central reaction versus non-central reaction.

\begin{figure}[htb]
\centering
\includegraphics[width=0.8\linewidth,angle=0,clip=true]{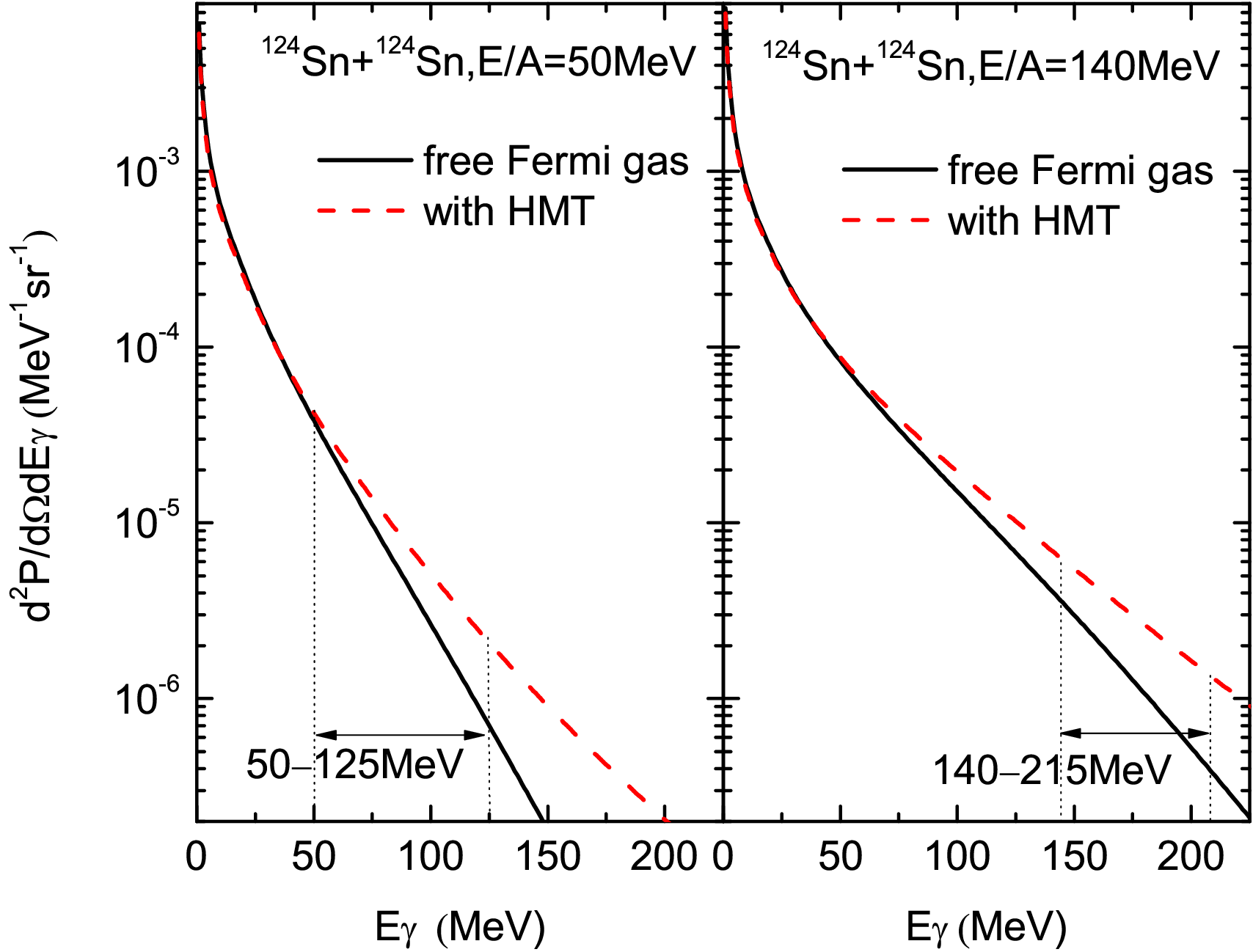}
\caption{The double differential probabilities of photon production in ${}^{124}$Sn + ${}^{124}$Sn central reactions at beam energies of 50 MeV/nucleon and 140 MeV/nucleon, respectively.}
\end{figure}
To examine the effect of HMT on the energy distribution of the bremsstrahlung photons, we plot the double differential probabilities of photon production as a function of photon energy from the head-on collisions of ${}^{124}$Sn + ${}^{124}$Sn with FFG and HMT at beam energies of 50 MeV/nucleon (left window) and 140 MeV/nucleon (right window) in Fig. 4, where the SBKD interaction is employed. It is clearly seen from the energy spectra of bremsstrahlung photons that the emissions of high energy photons are very sensitive to the HMT at both the two beam energies. It is also interesting to see that the effect of HMT on the double differential probability of photon production increases with the increasing photon energy $E_{\gamma}$. For example, at incident beam energy of 50 MeV/nucleon, the multiplicity of hard photons within energy range 50 MeV $\leq E_{\gamma} \leq$ 125 MeV increases from $9.0\times10^{-3}$ in FFG case to $1.2\times10^{-2}$ in HMT case. This corresponds to a 33\% growth of the high-energy photon production. And at 140 MeV/nucleon, the multiplicity of hard photons with energy 140 MeV $\leq E_{\gamma} \leq$ 215 MeV increases from $1.45\times10^{-3}$ in FFG case to $3.0\times10^{-3}$ in HMT case. The reason why the production of hard photons in the HMT case is higher than that of FFG case is because the HMT created by n-p SRC increases kinetic energies of a certain proportion of protons and neutrons in nuclei, thus more high energy photons are emited.

\begin{figure}[htb]
\centering
\includegraphics[width=0.8\linewidth,angle=0,clip=true]{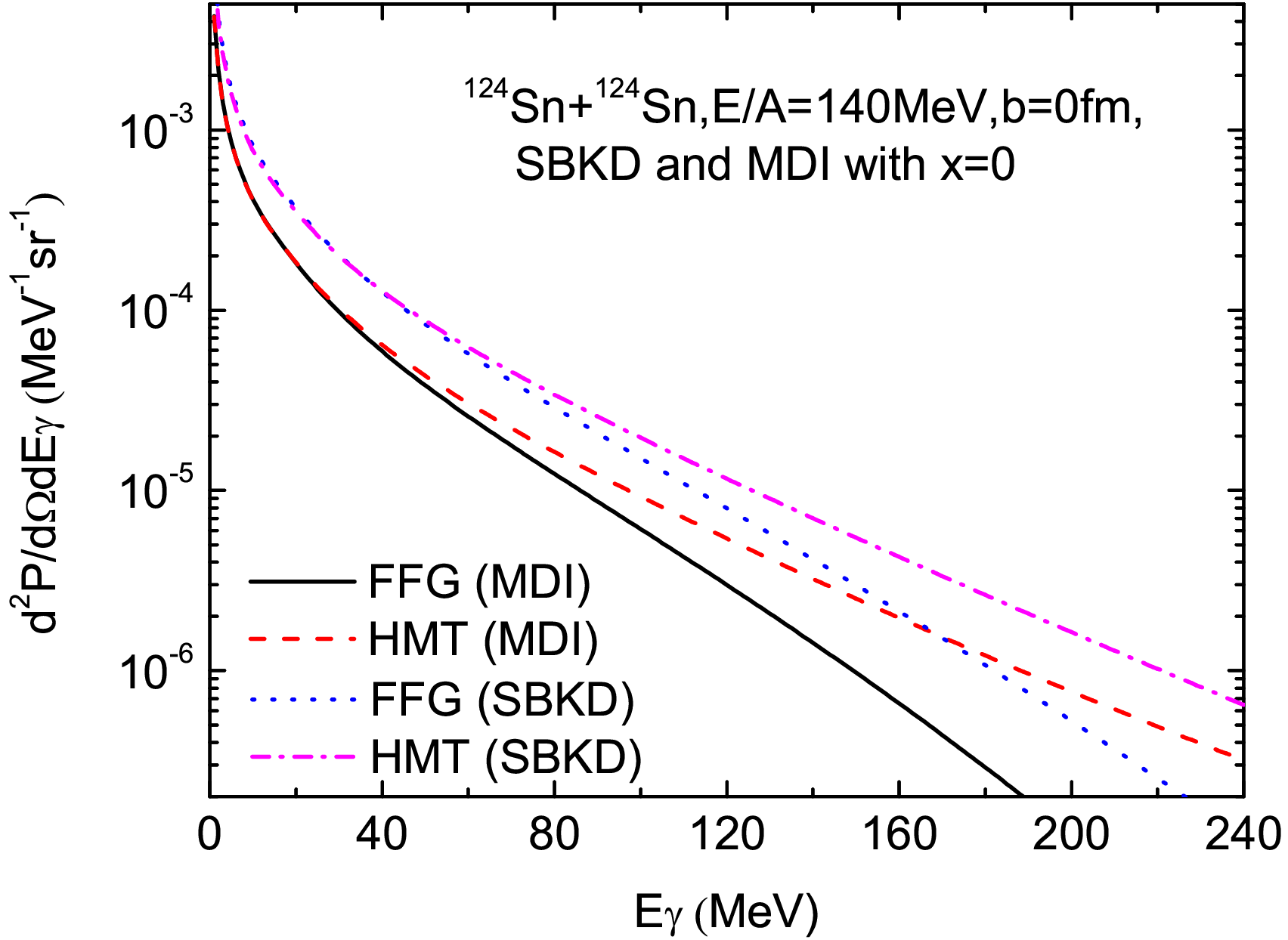}
\caption{The double differential probability of bremsstrahlung photons in ${}^{124}$Sn + ${}^{124}$Sn central reactions at a beam energy of 140 MeV/nucleon using SKBD and MDI interactions, respectively.}
\end{figure}
The simulation results discussed above are obtained by using the SBKD mean-field potential in IBUU model. To investigate the influence of different mean-field potentials on the photon emission in intermediate energy heavy-ion collisions in both FFG and HMT cases, we further compare the results of ${}^{124}$Sn + ${}^{124}$Sn reactions with SBKD \cite{42} and MDI interactions \cite{43}. Fig. 5 shows the double differential probability of photon production in the central ${}^{124}$Sn + ${}^{124}$Sn reactions at a beam energy of 140 MeV/nucleon using SBKD and MDI (with $x=0$) interactions, respectively. As expected, it is clearly seen that the effect of HMT in ${}^{124}$Sn + ${}^{124}$Sn reaction with MDI interaction is still obvious on the production of high-energy photons. We also see that the difference of the effect of HMT on bremsstrahlung photon production is relatively obvious between the two mean-field potentials. The emission of bremsstrahlung photons from the reaction with SBKD interaction is higher than that with MDI interaction in both FFG and HMT cases.

\begin{figure*}[htb]
\centering
\includegraphics[width=1.0\linewidth,angle=0,clip=true]{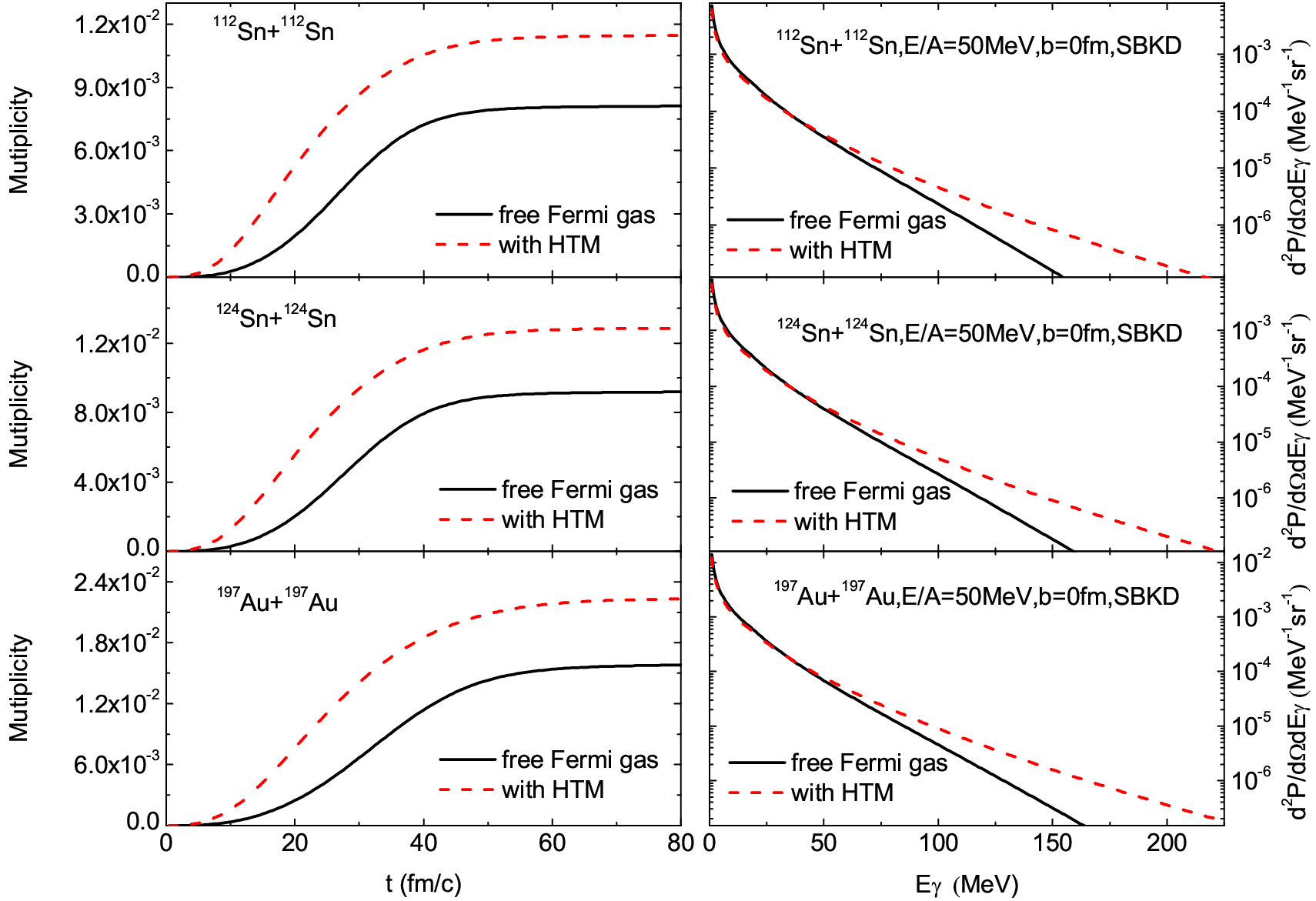}
\caption{The time evolution of photon multiplicity with energy 50 MeV $\leq E_{\gamma} \leq$ 275 MeV and the double differential probability of photon production as a function of photon energy in the central reactions of ${}^{112}$Sn + ${}^{112}$Sn, ${}^{124}$Sn + ${}^{124}$Sn and ${}^{197}$Au + ${}^{197}$Au using SBKD interaction at a beam energy of 50 MeV/nucleon with FFG and HMT.}
\end{figure*}
We also study the effect of HMT using the emission of bremsstrahlung photons for more neutron-rich systems. Fig. 6 gives the time evolution of photon multiplicity with energy 50 MeV $\leq E_{\gamma} \leq$ 275 MeV and the double differential probability of photon production as a function of photon energy in the central reactions of ${}^{112}$Sn + ${}^{112}$Sn, ${}^{124}$Sn + ${}^{124}$Sn and ${}^{197}$Au + ${}^{197}$Au using SBKD interaction at a beam energy of 50 MeV/nucleon, in the FFG and HMT cases, respectively. We see that the photon productions, especially the high energy photons are obviously sensitive to the HMT for all the three systems. Such as for the ${}^{197}$Au + ${}^{197}$Au reaction, the multiplicity of hard photons increases from $1.6\times10^{-2}$ in FFG case to $2.23\times10^{-2}$ in HMT case. It is also seen that the photon productions are increased with the growth of nucleon numbers from $^{112}$Sn to $^{197}$Au. For example, the photon multiplicity in the FFG case increases from $8.1\times10^{-3}$ in the ${}^{112}$Sn + ${}^{112}$Sn reaction to $9.2\times10^{-3}$ in the ${}^{124}$Sn + ${}^{124}$Sn reaction.

We have shown above that the bremsstrahlung photons are very sensitive to the high-momentum component in nucleon momentum distribution of neutron-rich nuclei.
However, there are several uncertainties such as the nucleon-nucleon (NN) scattering cross section and photon production probability, which may affect the photon emissions \cite{32,33}.
In the ref. \cite{32}, the uncertainties have been discussed and the ratio of differential photon production is proposed to reduce these unwanted factors as well as the systematic error.
We here use the same method and define the ratio of double differential photon production from two reactions as follows:
\begin{equation}
R_{p}=\frac{d^{2}P/d\Omega dE_{\gamma}\,({}^{124}Sn+{}^{124}Sn,50\ \mathrm{MeV/nucleon})}{d^{2}P/d\Omega dE_{\gamma}\,({}^{124}Sn+{}^{124}Sn,140\ \mathrm{MeV/nucleon})},\tag{11}
\end{equation}

\begin{figure}[htb]
\centering
\includegraphics[width=0.8\linewidth,angle=0,clip=true]{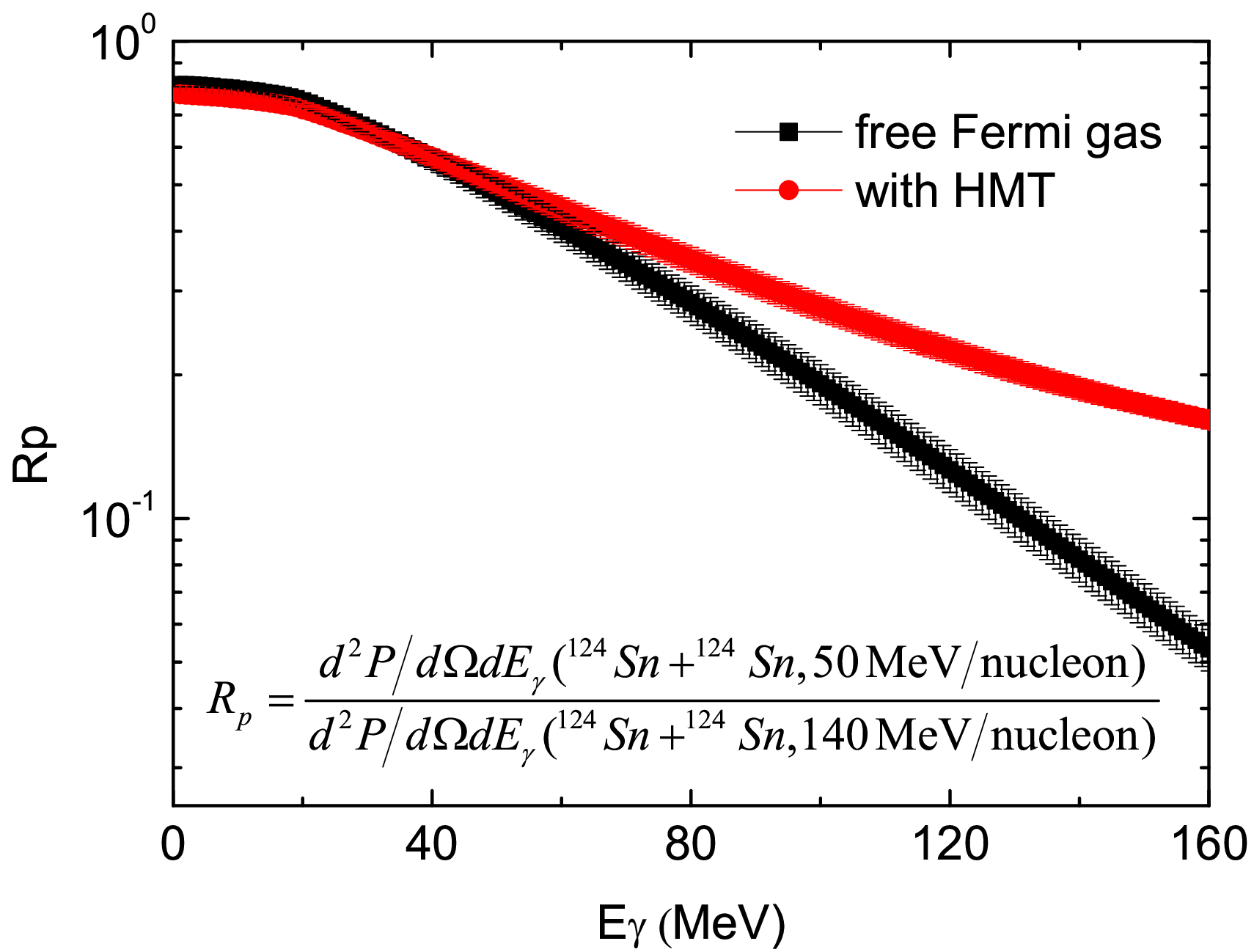}
\caption{Ratio of double differential photon production probabilities with error bar in central reactions of ${}^{124}$Sn + ${}^{124}$Sn at beam energies of 50 MeV/nucleon and 140 MeV/nucleon.}
\end{figure}
Fig. 7 shows the ratio of double differential photon production probability $R_p$ with the error bar as a function of photon energy $E_{\gamma}$ in the central reactions of ${}^{124}$Sn + ${}^{124}$Sn at beam energies of 50 MeV/nucleon and 140 MeV/nucleon.
We notice that the value of $R_p$ is less than 1, which is natural since the higher incident beam energy can lead to larger photon yield.
It is also clear that $R_p$ for energetic photon is quite sensitive to the HMT. The HMT can lead to an obvious increase of $R_p$ compared with FFG case in high energy photon region.
The difference of $R_p$ between HMT and FFG increases with photon energy $E_{\gamma}$ up to about 10.5\%.
This is helpful to distinguish the HMT experimentally. Thus the ratio $R_p$ can be a possible probe of the high-momentum component in nucleon momentum distribution of nuclei.
We also see that the error bars of $R_p$ in the HMT case become much smaller than those in the FFG case with the increasing of photon energy $E_{\gamma}$. This is possibly due to the fact that the higher energy photons are mainly produced from the collisions of nucleons with higher momentum, then the uncertainties of photon production probability and NN cross section become relatively weaker when taking the HMT in nucleon momentum distribution into consideration.

\section{summary}
In summary, we have carried out a study about the effect of HMT in nucleon momentum distribution of neutron-rich nuclei $^{112}$Sn, $^{124}$Sn and $^{197}$Au on the emission of bremsstrahlung photons from intermediate energy heavy-ion reactions based on the IBUU transport model. We found the proton-neutron bremsstrahlung photons are sensitive to the high-momentum component in nucleon momentum distribution, and the HMT can lead to an obvious increase of  hard photon production in the reactions of neutron-rich nuclei. With the growth of incident beam energy, the photon production is promoted in both HMT and FFG cases, while the effect of HMT on the bremsstrahlung photons becomes weaker at a higher incident beam. From the double differential probability of photon production, we show that the difference about the effect of HMT on the bremsstrahlung photon production is relatively obvious between the SBKD and MDI mean-field potentials.
We also investigate the effect of HMT on the photon emission in different systems such as $^{112}$Sn, $^{124}$Sn and $^{197}$Au. It is found the effect of HMT is universal in all the systems and
the production of bremsstrahlung photons is increased from the reaction of ${}^{112}$Sn + ${}^{112}$Sn to ${}^{197}$Au + ${}^{197}$Au.
To reduce the uncertainties such as the photon production probability and NN scattering cross section as well as the systematic error, the ratio of double differential photon production probability for the reactions with different beam energies $R_p$ is proposed as possible probe to study the HMT in nucleon momentum distribution of nuclei.
\\
\\
\section*{ACKNOWLEDGMENTS}
The work is supported by Guizhou Provincial Science and Technology Projects (ZK[2022]029).

\end{document}